\documentclass[aps,prl,floatfix,showpacs,showkeys,twocolumn,superscriptaddress]{revtex4-1}
\usepackage{bm}
\input{epsf}

\begin{document}

\title{Formation of the Bielectron in a 2D System due to Spin-Orbit Interaction and Image Forces}

\author{M.M. Mahmoodian}
 \affiliation{Rzhanov Institute of Semiconductor Physics, Siberian Branch of the Russian Academy of Sciences, \\Novosibirsk, 630090, Russia}
 \affiliation{Novosibirsk State University, Novosibirsk, 630090, Russia}

\author{A.V. Chaplik}
 \affiliation{Rzhanov Institute of Semiconductor Physics, Siberian Branch of the Russian Academy of Sciences, \\Novosibirsk, 630090, Russia}
 \affiliation{Novosibirsk State University, Novosibirsk, 630090, Russia}

\date{\today}

\begin{abstract}
It is shown that two electrons located in a quantum well near a metal electrode attract each other due to the spin-orbit interaction (SOI) of the Bychkov-Rashba type and the electrostatic image forces. Using the example of a simple model, it is shown that, with quite attainable values of the characteristic parameters of the system, the effective attraction caused by SOI prevails over the Coulomb repulsion, and the formation of a bielectron becomes possible.
\end{abstract}

\maketitle

Consider a quantum well (QW) in dielectric half-space $z<0$ while the upper half-space $z>0$ is occupied by an ideal metal (see Fig.~1). Each of the two electrons in the well interacts directly with the other one and with image charges. In any 2D system with asymmetry in the growth direction there exists the spin-orbit interaction (SOI) of the Bychkov-Rashba type \cite{b1}: $V_{SO}=\alpha{\bf n}[\bm{\sigma}{\bf p}]$, where $\alpha$ is the SOI constant, ${\bf n}$ is the unit normal vector, $\bm{\sigma}$ are Pauli matrices, and ${\bf p}$ is the 2D momentum.
\begin{figure}[h]\label{fig1}
\leavevmode\centering{\epsfxsize=7cm\epsfbox{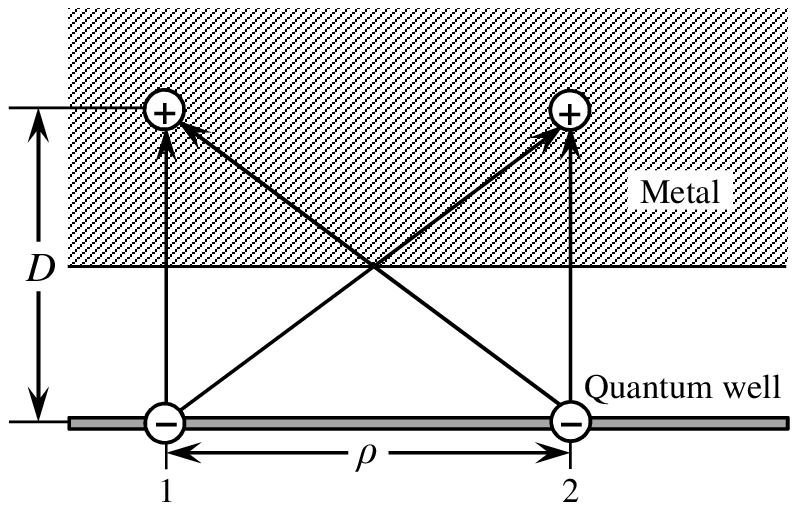}}
\caption{Fig.~1. Sketch of the structure. The arrows show the directions of the forces acting on the electrons.}
\end{figure}

In the situation of the Fig.~1 the asymmetry is provided by the metal electrode. As is known, the coefficient $\alpha$ depends on the $z$-component of the electric field $F_z$ experienced by electrons \cite{b2,b3}. In the problem in question the latter consists of two parts: 1) interaction with its own image and 2) interaction with the image of neighboring electron. It is shown in \cite{b2,b3} that $|\alpha|$ increases with increasing $F_z$. In the two electron system the total field $F_z$  depends on the interparticle separation $\rho$ and increases when $\rho$ decreases:
\begin{eqnarray}\label{f1}
F_z=\frac{e}{\varepsilon D^2}+\frac{eD}{\varepsilon\left(D^2+\rho^2\right)^{3/2}},~~~\rho=\left|\bm{\rho}_1-\bm{\rho}_2\right|.
\end{eqnarray}
Here $\varepsilon$ is the dielectric permeability, $D/2$ is the distance between QW and metal.

The lower branch of the SOI-split energy spectrum of electrons has the negative minimal value $-m\alpha^2/2$ when the 2D momentum corresponds to the extrema loop. Decrease of $\rho$ results in increase of $\alpha$ and, hence, the minimum of the energy becomes deeper. Thus, the energy of the two-electron system lowers when the interparticle separation decreases due to SOI and image forces. Simultaneously the Coulomb repulsion raises the energy and, if the balance turn out negative, the bound state of two electrons -bielectron- becomes possible. Evidently the corresponding energy level must be lower than ($-m\alpha^2/2$) - the sum of minimal energies of the two electrons at $\rho\rightarrow\infty$.

\subsection*{Quantitative consideration.}

In our problem the total electron field ''seen'' by each electron has both the normal component $F_z$ and the lateral one ${\bf F}_\tau$. As we will show in the last part of this paper the ${\bf F}_\tau$ contribution may be disregarded at least in the principal question: whether the bound state exists at all? With accounting for only $F_z$ the Hamiltonian reads:
\begin{eqnarray}\label{f2}
\hat{H}=\hat{H}_0+A\left\{F_z(\rho)\left(\left[\hat{\bf{p}}_1\bm{\sigma}_1\right]+\left[\hat{\bf{p}}_2\bm{\sigma}_2\right]\right){\bf n}\right\},
\end{eqnarray}
\begin{eqnarray}\label{f3}
\hat{H}_0=\frac{\hat{\bf{p}}_1^2+\hat{\bf{p}}_2^2}{2m}+\frac{2e^2}{\varepsilon}\left(\frac{1}{\rho}-\frac{1}{\sqrt{\rho^2+D^2}}\right),
\end{eqnarray}
where $A$ is determined by the field dependence of the SOI constant $\alpha=AF_z$, ${\bf n}$ is the unit vector along $z$-axis, $\left\{\hat{Q}\hat{T}\right\}=\frac12\left(\hat{Q}\hat{T}+\hat{T}\hat{Q}\right)$. Two particle wave function is the bi-spinor $(\psi_1\psi_2\psi_3\psi_4)$. Four equations for $\psi_i$ (i=1,2,3,4) have to be solved with the evident boundary conditions: each $\psi_i$ is regular at $\rho=0$ and tends to zero when $\rho\rightarrow\infty$. The details of calculations are given in our extended paper submitted to JETP Letters. Here we report the main findings.

Any solutions of the Schr\"{o}dinger equation for the Hamiltonian (\ref{f2}) takes the form $e^{i{\bf PR}}\psi(\bm{\rho})$, where ${\bf R}=(\bm{\rho}_1+\bm{\rho}_2)/2$, $\bm{\rho}=\bm{\rho}_1-\bm{\rho}_2$. For ${\bf P}=0$ the ground state bi-spinor is
\begin{eqnarray}
\psi_1=\chi_1(\rho)e^{-i\varphi},~\psi_2=\chi_2(\rho),~\psi_3=\chi_3(\rho),~\psi_4=\chi_4(\rho)e^{i\varphi},\nonumber
\end{eqnarray}
where $\rho$ and $\varphi$ are the polar coordinates of the vector $\bm{\rho}$. The bound state is described by the solution in which $\chi_1=\chi_4, \chi_2=-\chi_3$. Such a solution corresponds to the total spin $S=1$ and provides the permutative anti-symmetry of the total wave function (at permutation $\varphi$ goes over to $\varphi+\pi$, $\psi_1$ and $\psi_4$ change signs, $\chi_2$ goes over to $\chi_3$ and vice versa while $\chi_2=-\chi_3$). Thus, the pairing (if it exists!) occurs in the triplet state.

We consider a simplified exactly solvable model where all the coefficients $\chi_i(\rho)$ depending on $\rho$ in the equations for $\chi_1$, $\chi_2$ are replaced by rectangular barriers $V_{0i}\theta(L-\rho)$, $\theta$ is Heaviside step-function, $L$ is a free parameter, $V_{0i}$ is taken from the relation $\pi L^2V_{0i}=\int V_i(\rho)\rho d\rho d\varphi$. With boundary conditions of continuity of $\chi_1$, $\chi_2$ and their derivatives at $\rho=L$ we come to the $4\times4$ determinant containing Bessel functions $J_{0,1}$ and $K_{0,1}$. Zeros of the determinant define the energy spectrum of the system. Just to demonstrate the possibility of principle to form the bound state we used the direct variational approach with a simple trial function. For the bielectron energy level $E_0$ we found
\begin{eqnarray}\label{f4}
E_0=-m\alpha_0^2\left(1+\frac{2D^2}{L^2}\right)^2+\frac{4e^2D}{\varepsilon L^2},
\end{eqnarray}
with $\alpha_0=Ae/\varepsilon D^2$ - SOI-constant for a single electron.

\subsection*{Numerical example.}

From the data given in \cite{b2,b3} one can find that in quantum wells of $Bi_2Se_3$ the coefficient $A$ equals $2\cdot10^2~\mbox{cm}^2/(\mbox{V}\cdot\mbox{s})$. Take $L=3D$, $\varepsilon=10$, $m=0.1m_e$, $D=3$~nm, then $\alpha_0=2.5\cdot10^7~\mbox{cm}/\mbox{s}$ and for the bielectron bound energy $\Delta=|E_0+m\alpha_0^2|$ we get $\Delta\simeq5.4~\mbox{meV}$, the size of the bound state $r_0\approx4$~nm.

\subsection*{Acknowledgements.}

We thank I.F. Ginzburg, V.A. Volkov, M.V. Entin and V.M. Kovalev for useful comments and discussion of our work. Financial support of RFBR (Grant No.16-02-00565) and Program of RAS (Project 0306 - 2018-0007) is gratefully acknowledged.

\end{document}